\documentclass[notitlepage,aps,pra,twoside,11pt,tightenlines, superscriptaddress,showkeys,nofootinbib]{revtex4-1}
\usepackage{graphicx}
\usepackage{amsmath,amsfonts,amssymb,amsthm,amsbsy,mathtools}
\usepackage{bbold}
\usepackage{mathrsfs}

\newcommand{\ket}[1]{\mathop{\left|#1\right>}\nolimits}            

\newcommand{\brk}[2]{\langle #1 | #2 \rangle}
\newcommand{\kbr}[2]{| #1\rangle\!\langle #2 |}
\newcommand{\dx}[1]{\mathrm{\,d} #1}             
\def\bk{\mathbf{k}}
\def\df{\overset{\rm def}{=}}

\def\bbR{\mathbb{R}}
\def\bbC{\mathbb{C}}

\newcommand{\nn}{\nonumber}

\newcommand{\Tr}[1]{\mathop{{\mathrm{Tr}}_{#1}}}            
\DeclareMathOperator\erf{erf}
\DeclareMathOperator\atanh{arctanh}

\def\a{\alpha}
\def\b{\beta}
\def\g{\gamma}

\def\vr{\varrho}

\def\s{\sigma}

\def\la{\lambda}
\def\La{\Lambda}

\def\vt{\vartheta}
\def\t{\vartheta}

\def\D{\mathcal{D}}
\def\L{\mathcal{L}}

\def\P{\mathcal{P}}

\newmuskip\pFqmuskip   

\newcommand*\pFq[6][8]{%
  \begingroup 
  \pFqmuskip=#1mu\relax
  \mathcode`\,=\string"8000
  \begingroup\lccode`\~=`\,
  \lowercase{\endgroup\let~}\pFqcomma
  {}_{#2}F_{#3}{\left[\genfrac..{0pt}{}{#4}{#5};#6\right]}
  \endgroup
}
\newcommand{\pFqcomma}{\mskip\pFqmuskip}

\begin{document}

\title{Quantum and classical capacity boosted by a Lorentz transformation}

\author{Kamil Br\'adler}
\email{kbradler@ap.smu.ca}
\affiliation{Department of Astronomy and Physics,
    Saint Mary's University,
    Halifax, B3H 3C3, Canada}

\author{Esteban Castro-Ruiz}
\author{Eduardo Nahmad-Achar}
\affiliation{Instituto de Ciencias Nucleares,
    Universidad Nacional Aut\'onoma de M\'exico, Mexico D. F., Mexico}

\begin{abstract}
    In this paper we show that the quantum channel between two inertial observers who transmit quantum information by sending realistic photonic wave packets is a well-studied channel in quantum Shannon theory -- the Pauli channel. The parameters of the Pauli channel and therefore its classical and quantum capacity depend on the magnitude of the Lorentz boost  relating the two observers.
    The most striking consequence is that two inertial observers  whose Pauli channel   has initially zero quantum capacity can achieve nonzero quantum communication rates (reaching in principle its maximal value equal to one) by applying a boost in the right direction. This points to a fundamental connection between quantum channel capacities and special relativity.
\end{abstract}



\maketitle

Photons carrying quantum information encoded in the polarization degrees of freedom between two inertial observers were first studied in~\cite{alsingmilburn}. Photons were considered to be momentum and helicity eigenstates but it soon became clear that a more realistic description is given by localized wave packets~\cite{czachor,PT,Gingrich}. This opened a Pandora's box where among the most pressing problems is the definition of a polarization density matrix. It turns out that, apart from specially crafted wave packets~\cite{caban}, there is no covariant definition of a polarization (helicity) density matrix. The main reason lies in the nontrivial dependence of the helicity on the momentum~\cite{wigner}. One of the consequences is the impossibility to trace over the momentum degree of freedom, leading to some interesting effects for two inertial observers trying to communicate by sending such wave packets~\cite{PT}.

Here we approach the problem from an entirely different perspective. When it comes to quantum communication over a noisy quantum channel, the important quantity is the classical or quantum channel capacity studied in quantum Shannon theory~\cite{wilde} (or \cite{holevo} for more mathematically  oriented readers). Quantum channel capacities quantify the highest achievable rate at which nearly perfect transmission of quantum~\cite{QC} or classical~\cite{holevo} messages  through a noisy quantum channel is possible. Quantum codes prepared for this purpose by the sender  can be used for transmission of classical~\cite{holevo} or quantum information~\cite{QC}.
In order to establish the channel capacity, it is necessary to identify the quantum channel first. This is done by careful study of the physical scenario. In our case we consider two inertial observers where, without loss of generality, one of them is considered to be at rest and the other is moving at a constant relativistic speed. Under these conditions, the most suitable carriers of information seem to be photons with helicity (circular polarization) degrees of freedom. Our intention is to analyze a realistic scenario where the photonic states are spatially localized polychromatic wave packets whose momentum distribution is a reasonably chosen square-integrable function. We do not rely on less realistic schemes with momentum/helicity eigenstates~\cite{alsingmilburn}  or linearly polarized wave packets~\cite{caban}. The first step the sender must take is to map a logical qubit $\psi=\a\ket{0}+\b\ket{1}$ to a sufficiently realistic wave packet~$\Psi_0$. The wave packet is then Lorentz transformed to the receiver's frame. An important ingredient is therefore detection. We use the highly realistic and simple detection mechanism proposed in~\cite{Aiello}. As a consequence, the two main reasons that jointly contribute to the appearance of a noisy channel are (i) the Lorentz transformation itself and (ii) the detection process. Importantly,  even the sender in his own reference frame cannot simply undo the mapping $\psi\mapsto\Psi_0$. The reason is the intentionally low level of sophistication of the detection process~\cite{Aiello}.

We identify the induced quantum channel to be a Pauli channel whose parameters are functions of the boost and the  wave packet variance. We calculate the classical~\cite{king} and quantum~\cite{smithPauli} capacities of the channel. More precisely, the quantum capacity of a general Pauli channel is not known to possess a calculable formula, but a lower bound on reliable quantum communication is known (the hashing bound~\cite{hash}), and an upper bound on the zero quantum capacity based on a no-cloning argument is known as well~\cite{cerfPauli}.   As one of the consequences we conclude that for two observers whose quantum capacity is initially zero (due to a poorly prepared wave packet), it can be increased arbitrarily close to its maximum value by a boost in the right direction. The two observers can be initially at rest or moving with respect to each other. The amplification effect due to a Lorentz boost exists for the classical capacity as well. This inexorably points to a deep connection between quantum Shannon theory and special relativity similar to that in classical Shannon theory~\cite{jarett}. We also clarify the reported occurrence of a non-completely positive map in a similar situation~\cite{PT} and show that, despite its validity, it actually plays no role in realistic quantum communication between two inertial observers.

The communication setup consists of three steps. The sender first maps his logical qubit to a photonic wave packet: $\psi\mapsto\Psi_0$. The general form of $\Psi_0$ reads~\cite{tung} $\ket{\Psi_0}=\sum_{\la=\pm}\int_{\bbR^3} f_\la(\bk)\ket{k,\la}\dx{\mu}(k)$, where $\dx{\mu}(k) = \dx^3{k}/[(2\pi)^32 k^0]$ is the Lorentz invariant integration measure and $k=(k^0,\bk)$ is the 4-momentum vector.  We denote $f_+({\bk})=\a f({\bk})$ and $f_-({\bk})=\b f({\bk})$ and choose $f({\bk})$ to be a Gaussian momentum distribution with an axial symmetry.

The second step is the Lorentz transformation of the wave packet. In this paper we focus on a Lorentz boost $\La=B_z(\zeta)$ where $\zeta=\atanh{v_z}$ is the rapidity (assuming $c=1$) and $-1<v_z<1$ is the velocity. The induced unitary transformation of the wave packet is denoted by $U(\La)$ and its action $U(\La)\Psi_0=\Psi_\zeta$ results in the modification of the envelope function $f(\La^{-1}\bk)$. At this point we note that for realistic wave packets it is natural to assume that the momentum variance in the propagation direction is much smaller than the radial variance ($\s_z\ll\s$). This approximation gives rise to the momentum distribution function used in this paper
\begin{equation}\label{eq:approxEnvelop}
  |f(\La^{-1}\bk)|^2 = {1\over N} {\exp\left(-\frac{ \sin^2\vt}{\Gamma^2(\sinh\zeta+\cosh\zeta\cos\vt)^2 }\right) \over\sinh{\zeta}+\cosh{\zeta}\cos{\vt}},
\end{equation}
where $\Gamma$ is the wave packet spread, $\vt$ is the polar angle of $\bk$ and $N$ is chosen such that the covariant normalization condition $\int_{\bbR^3} |f(\mathbf{k})|^2 \dx{\mu}(k)=1$ is satisfied. The derivation of (\ref{eq:approxEnvelop}) is presented in Appendix \ref{ap:wavpackconstr} in great detail. We emphasize that the approximation is valid for all $\zeta\in\bbR$: If we take the limit $\sigma_z \to 0$ and then Lorentz transform, the result is identical to Lorentz-transforming the wave packet with finite $\sigma_z$ and then taking the limit $\sigma_z \to 0$. It is in this sense that our wave packet has a well-defined transformation (cf. Appendix \ref{ap:wavpackconstr}).

The final step is the recovery of the information encoded in the helicity degree of freedom of the wave packet, leading to the desired output density matrix: $\Psi_\zeta\mapsto\vr_\zeta$. It is far from obvious how to achieve this goal because the momentum and helicity degrees of freedom are not independent. A simple partial trace over the momenta is not a correct description of helicity states~\cite{PT} because the helicity Hilbert space (Wigner's `little' space; see Appendix \ref{ap:wavpackconstr}) can be thought of as a fiber of a coset space (the positive light-cone minus the origin)~\cite{dutta,sexl} and so each $k$ the  wave packet is constructed from `carries' its own Hilbert space $\bbC^2_k$. However, we can define an effective polarization density matrix $\vr_\zeta$ from the expected values of measurements on the complete state $\Psi_\zeta$. The  density matrix formed in this way contains all the information regarding possible polarization measurements. We follow the construction given by~\cite{Aiello}, where the detection is represented by
\begin{equation}\label{eq:W}
W(\bk) =
\begin{bmatrix}
\frac{\cos\phi\cos{\vt}}{\sqrt{1- \cos^2\phi\sin^2{\vt}}} && \frac{-\sin\phi}{\sqrt{1- \cos^2\phi\sin^2{\vt}}} \\
\frac{\sin\phi\cos{\vt}}{\sqrt{1- \sin^2\phi\sin^2{\vt}}} && \frac{\cos\phi}{\sqrt{1- \sin^2\phi \sin^2{\vt}}}
\end{bmatrix}.
\end{equation}
What is the meaning of this transformation? Consider a linear polarizer whose axis is $\mathbf{\hat{z}}$. Then $W(\bk)$ is a transmission matrix whose elements are $w_{ij}=\brk{k,i}{k,j}$, where $i=\{0,1\}=\{\mathbf{\hat{x}},\mathbf{\hat{y}}\}$ and $j=\{0,1\}=\{H,V\}$. The state $\ket{k,\mathbf{\hat{x}}}\ (\ket{k,\mathbf{\hat{y}}})$ is defined as the state behind the device that is oriented along $\mathbf{\hat{x}}\  (\mathbf{\hat{y}})$. Given  $W(\bk)$ for every $\bk$, the output polarization density matrix $\vr_\zeta$ is formed by averaging the polarization vectors over all momenta. Note that $W(\bk)$ is neither a projector (or POVM) nor a unitary matrix (except for $\vt=0$, where it behaves as a polarization rotator). This is crucial because any form of measurement would destroy quantum information and lead to zero quantum capacity.
\begin{figure}[h]
    \includegraphics[scale=1]{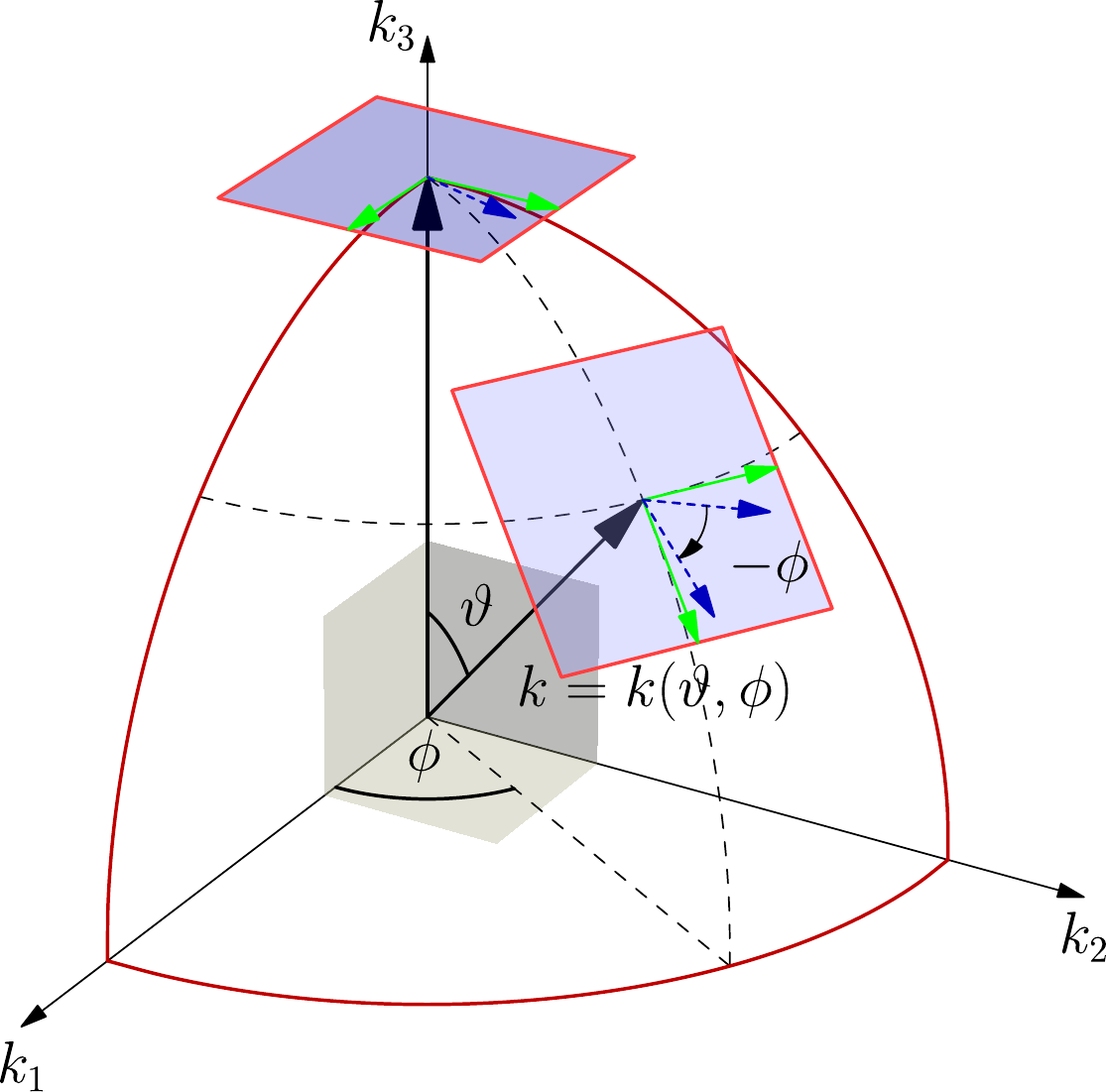}
    \caption{(Color online) The general wave packet can be thought of as a collection of $k$ vectors modulated by an envelope function. Each vector has a (complex) helicity space attached whose real projection can be visualized as an $\bbR^2$ plane perpendicular to the momentum vector (represented by two tangential planes). The amplitudes $a$ and $b$ of the real helicity vectors given by the projection of the blue dashed arrows on the green helicity basis  vectors lying in the plane  are identical for all $\bk$'s, but because the momentum vectors point in different directions in momentum space, the  helicity vectors point in different directions in ambient $\bbR^3$ space (see the related discussion before Eqs.~(\ref{eq:W}) and~(\ref{eq:unrotation})). To correct for this effect, we `unrotate' the polarization vectors by applying Eq.~(\ref{eq:unrotation}). This compensates the rotation of the $\bk$ vector and is depicted in the plane perpendicular to the vector $k=k(\vt,\phi)$. But it works well only for small $\vt$. The unrotation becomes less effective as $\vt$ increases, and for $\vt\to\pi/2$ it is useless since the helicity vector points `downwards'. The important point is that, without the unrotation, reliable quantum communication is impossible (see Eq.~(\ref{eq:QCapUnrotation})).}
    \label{fig:unrotation}
\end{figure}

We can now put all the pieces together. Our task is to investigate the character and properties of the overall map $\P:\psi\mapsto\vr_\zeta$. The explicit form of $\vr_\zeta$ reads~\cite{Aiello}
\begin{equation}\label{eq:rhoOut}
  \vr_\zeta=\sum_{m,n=0,1}\kbr{m}{n}\int_{\bbR^3}|f(\La^{-1}\bk)|^2(aw_{m0}+bw_{m1})(\overline aw_{n0}+\overline bw_{n1})\dx{\mu}(k),
\end{equation}
where $a=(\a+\b)/\sqrt{2},b=i(\a-\b)/\sqrt{2}$, and the bar denotes complex conjugation (the reason behind this transformation is that we work in the helicity basis, whereas $W(\bk)$ is written in the horizontal/vertical polarization basis). We can imagine the wave packet as a collection of $\bk$ vectors each with a perpendicular plane $\bbR^2$ attached. The plane contains the real projections of the polarization vector $a\ket{k,H}+b\ket{k,V}$. In the ambient $\bbR^3$ space the real polarization projections for different $\bk$ point in different directions (see Fig.~\ref{fig:unrotation}). Consequently, such a wave packet is useless for sending quantum information, as shown in Eq.~(\ref{eq:QCapUnrotation}), given that our detection model is in terms of linear polarizers. This detection model (Eq.~(\ref{eq:W})) is simple, but sufficiently realistic. The intuitive explanation for it is that $W(\bk)$ essentially uniformly averages over polarization vectors for all $\bk$ and the coherence present in the off-diagonal elements of $\kbr{\psi}{\psi}$ is wiped out. Instead, we engineer our wave packet such that the coefficients in Eq.~(\ref{eq:rhoOut}) become
\begin{equation}\label{eq:unrotation}
  \begin{bmatrix}
    a \\
    b \\
  \end{bmatrix}
  \mapsto
  \begin{bmatrix}
    \cos{\phi} & \sin{\phi} \\
    -\sin{\phi} & \cos{\phi} \\
  \end{bmatrix}
    \begin{bmatrix}
    a \\
    b \\
  \end{bmatrix}.
\end{equation}
The transformation and its rationale are explained in Fig.~\ref{fig:unrotation}. The photon packet carries information whether or not Eq.~(\ref{eq:unrotation}) is used in the preparation of the state. When this {\it un-rotation} is not used, we cannot extract useful information from the reduced density operator constructed from our detection model, since, as mentioned above, although general and realistic in nature, it uses linear polarization measurements. The use of Eq.~(\ref{eq:unrotation}) in the preparation of the state, however, allows us to extract the codified information from such a reduced density matrix.

Remarkably, the sought-after map $\P$ turns out to be a Pauli channel, $\P:\vr\mapsto p_0\vr+\sum_{i=1}^3p_i\tau_i\vr\tau_i$, where $\tau_i$ are Pauli matrices (using the convention $\{1,2,3\}=\{x,y,z\}$) and $0\leq p_i\leq1$ satisfying  $\sum_{i=0}^3 p_i=1$. By suitably reparametrizing the input state $\psi$  using
$\a=\exp{(-i\chi)}\cos{(\xi/2)}$ and $\b=\exp{(i\chi)}\sin{(\xi/2)}$, the Pauli channel output reads
\begin{equation}\label{eq:PauliOutput}
  \P(\vr)={1\over2}{
    \begin{bmatrix}
        1+\la_3\cos{\chi}\sin{\xi} & \la_1\sin{\chi}\sin{\xi}-i\la_2\cos{\xi} \\
        \la_1\sin{\chi}\sin{\xi}+i\la_2\cos{\xi} & 1-\la_3\cos{\chi}\sin{\xi} \\
    \end{bmatrix}
    },
\end{equation}
where $\la_1=p_0+p_1-p_2-p_3,\la_2=p_0-p_1+p_2-p_3$ and $\la_3=p_0-p_1-p_2+p_3$.

The fact that our physical setup becomes a Pauli channel is highly nontrivial. It can be  seen by comparing the elements of the density matrix,~(\ref{eq:rhoOut}), derived in Appendix \ref{ap:wavpackconstr} (Eqs.~(\ref{eq:gFcns}) with Eq.~(\ref{eq:PauliOutput}). In addition, for this to be true, the following identities must be satisfied for $j=1,2$:
\begin{align}\label{eq:rhoOutCompsIdentities}
   \int_0^{2\pi}\int_0^{\vt_c} K(\vt,\zeta,\Gamma){g_j\over1-\cos^2{\phi}\sin^2{\t}}\dx{\phi}\dx{\t}=\int_0^{2\pi}\int_0^{\t_c} K(\t,\zeta,\Gamma){(-)^{j+1}g_{j+2}\over1-\sin^2{\phi}\sin^2{\t}}\dx{\phi}\dx{\t},
\end{align}
where $K(\t,\zeta,\Gamma)$ is the kernel from Eq.~(A.15) and $g_j$ are defined in Eqs.~(A.21). The validity of Eq.~(\ref{eq:rhoOutCompsIdentities}) is proved in Eq.~(A.23) and in the paragraph that follows. The origin of the upper integration bound $\t_c=\arccos{(-\tanh{\zeta})}$ is explained before Eq.~(A.16).

The main consequence is that we do not need to evaluate the density matrix integrals in~(\ref{eq:rhoOut}). It is not even desirable -- we are interested in the channel and its capacities as functions of $\zeta$ and $\Gamma$ (certainly not of $\chi$ or $\xi$!) and the above identification enables us to find  analytic and perturbative expressions (in $\Gamma$) for $\la_i$ (as illustrated for $\zeta=0$ in Eqs.~(B.3), (B.12) and~(B.4)).

\begin{figure}[t]
    \begin{center}
    \resizebox{10cm}{7cm}{\includegraphics{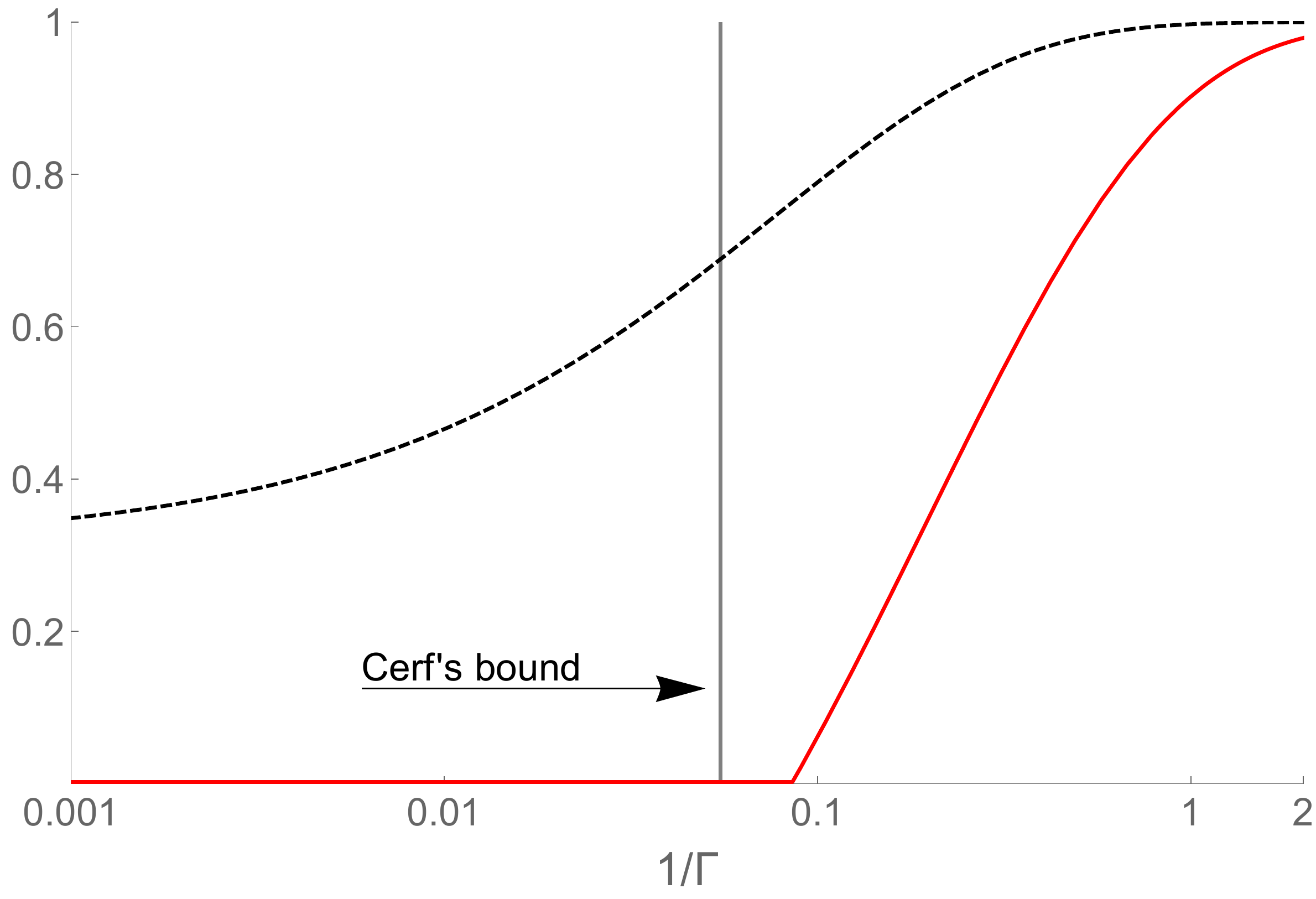}}
    \caption{The classical capacity $C(\P)$  in bits per channel (upper dashed curve) and a lower bound on the quantum capacity $Q_\uparrow(\P)$ in bits per channel (lower curve) of a Pauli channel plotted as a function of $1/\Gamma$. The zero lower bound $Q_\uparrow(\P)$  becomes true zero quantum capacity below Cerf's bound where $c_\downarrow^0\geq1/2$ (to the left of the vertical line).}
    \label{fig:caps}
    \end{center}
\end{figure}

Based on this insight, we are now ready to write down an  expression for the classical capacity and lower and upper bounds of the quantum capacity. Following the prescription given in~\cite{king} for general unital qubit channels, we write $C(\P)=1-H(x)$ for the classical capacity of a Pauli channel~$\P$, where $H(\{x,1-x\})=-x\log_2{x}-(1-x)\log_2{(1-x)}$ is the Shannon entropy, $x=(1+\max_i{|\la_i|})/2$ and $\max_i{|\la_i|}=\la_1$ in our case. For a lower bound on the quantum capacity we use the hashing (random coding) bound~\cite{hash} given by $Q_\uparrow(\P)=1-H(\{p_0,p_1,p_2,p_3\})$. On the other hand, the quantum capacity of a Pauli channel is zero if the following condition is satisfied~\cite{cerfPauli}: $c^0_{\downarrow}\df p_1+p_2+p_3+\sqrt{p_1p_2}+\sqrt{p_2p_3}+\sqrt{p_1p_3}\geq1/2$. This is an upper bound we draw our main conclusion from and together with $C(\P)$ and $Q_\uparrow(\P)$ it is plotted in Fig.~\ref{fig:caps}.

\begin{figure}[t]
    \begin{center}
    \resizebox{10cm}{7cm}{\includegraphics{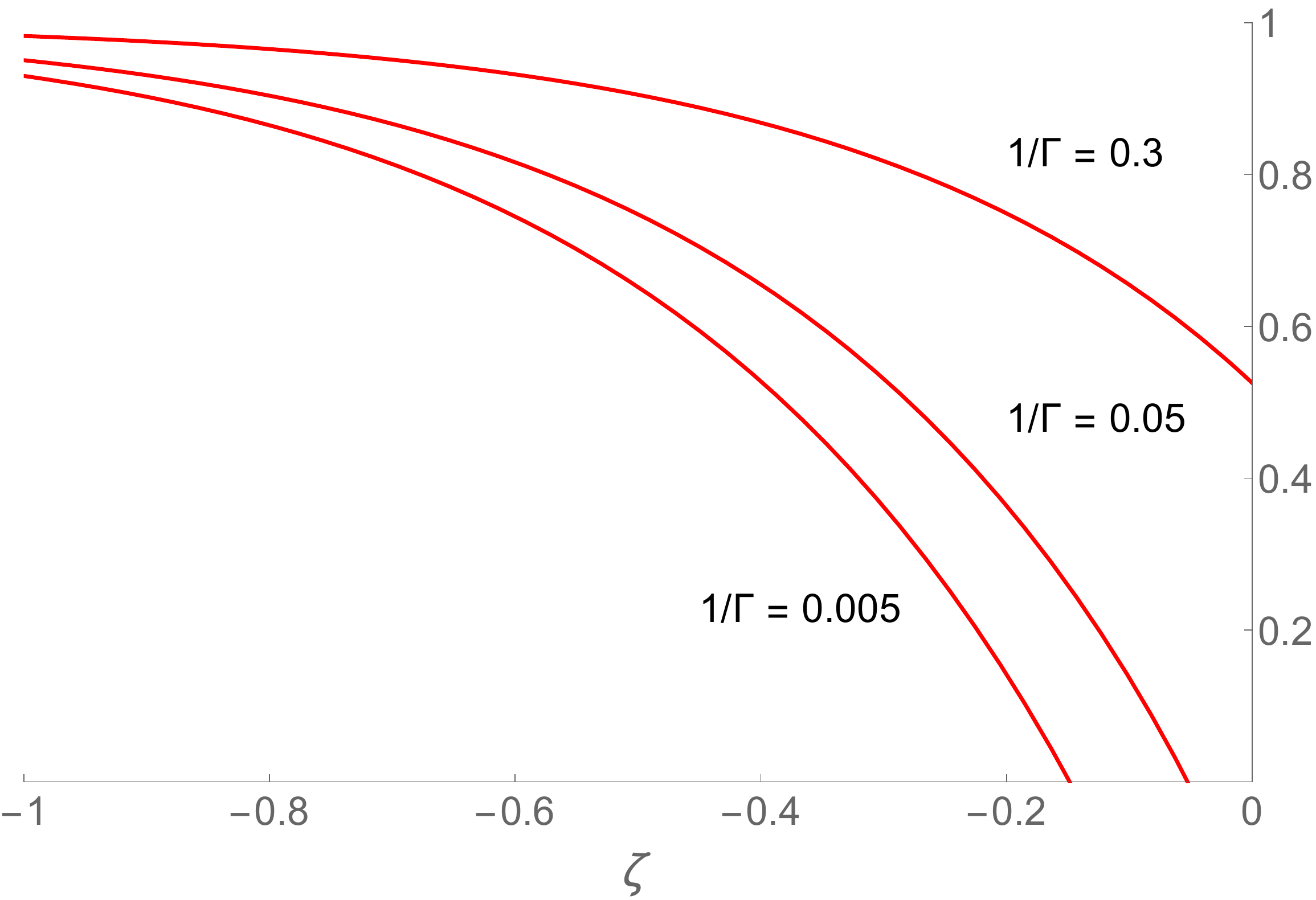}}
    \caption{The effect of a negative rapidity $\zeta$ in the $z$ direction (an approaching observer) on the quantum capacity's lower bound (in bits per channel) is illustrated for three different initial wave packets. Two wave packets whose quantum capacity is zero ($1/\Gamma=0.005$ and $1/\Gamma=0.05$, see Fig.~\ref{fig:caps}) can be boosted to nonzero values. Already for a nonzero value of quantum capacity (illustrated as $1/\Gamma=0.3$) the boost further increases the communication rate.}
    \label{fig:capsBoosted}
    \end{center}
\end{figure}

\begin{figure}[t]
   \resizebox{8cm}{!}{\includegraphics{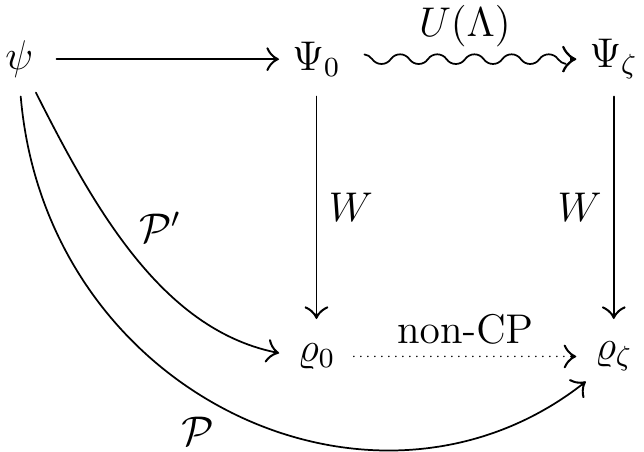}}
    \caption{Diagram describing the whole physical setup and the emergence of a relativistic Pauli channel $\P$ (note that $\P'$ is $\P$ for $\zeta=0$). For a detailed description see the paragraph preceding Appendix~\ref{ap:wavpackconstr}.}
    \label{fig:diagram}
\end{figure}

Assume an inertial observer who prepares a poor wave packet $\Psi_0$ whose spread $\Gamma$ leads to $c^0_{\downarrow}\geq1/2$. In this case, the quantum capacity is exactly zero\footnote{Note that another method useful to conclude that the quantum capacity is zero is to calculate whether the Pauli channel is entanglement-breaking~\cite{EB}. However, it turns out that our Pauli channels become entanglement-breaking deep inside Cerf's territory.} (recall that even the sender himself, equipped with our realistic detection  scheme, cannot reverse the mapping $\psi\mapsto\Psi_0$), and therefore no quantum communication is, in principle, possible (cf. Fig.~\ref{fig:caps}). If, however, one of the participants Lorentz boosts himself in such a way that the envelope function  Eq.~(\ref{eq:approxEnvelop}) becomes  sufficiently localized, then the hashing bound becomes strictly positive and reliable quantum communication is  possible. This is precisely what we see in Fig.~\ref{fig:capsBoosted} for negative rapidity $\zeta$, which means that the sender and the receiver are approaching each other. A similar increase as a consequence of the Lorentz boost is witnessed for the classical capacity (not depicted). Moreover, from the asymptotic behavior ($\Gamma\to0$ or $\zeta\to-\infty$) of the $\la_i$ functions it follows that both communication rates approach their maximal value one. This is because, in this limit, the envelope function becomes a delta function with all the helicity vectors aligned, turning the Pauli channel into a noiseless channel.

We therefore for the first time demonstrate an intricate connection between quantum Shannon theory and special relativity. This follows the footsteps of Ref.~\cite{jarett}, where the relation between classical Shannon theory and special relativity has been exposed. Several comments are in place. First, here we consider a Lorentz boost in the direction of propagation of the wave packet. A more general Lorentz transformation would lead to a more complicated behavior due to the presence of a nontrivial Wigner phase~\cite{Gingrich}. Second, there is a gap between Cerf's bound and the nonzero hashing bound (see Fig.~\ref{fig:caps}). The Pauli channel in this area is not a one-Pauli channel (defined as having any pair of $\{p_1,p_2,p_3\}$ zero) for which the hashing bound equals the quantum capacity itself~\cite{smithPauli}. Hence it may happen that the method based on highly degenerate quantum codes showing superadditivity of the optimized coherent information for Pauli channels~\cite{smithPauli} can lead to nonzero rates despite the hashing bound being zero. Third, if a wave packet is used without the polarization unrotation Eq.~(\ref{eq:unrotation}) (see Fig.~(\ref{fig:unrotation})), the resulting channel turns out to be $\D\circ\P_2$, where $\D$ is a qubit depolarizing channel~\cite{wilde} and $\P_2$ is a one-Pauli channel with $p_1=p_3=0$ and $p_0=p_2$. Because of the aforementioned property of the one-Pauli channels, we have $Q_\uparrow(\P_2)=Q(\P_2)=1-H(\{p_0,0,p_0,0\}=0$. By further using the bottleneck inequality for the quantum capacity we finally obtain
\begin{equation}\label{eq:QCapUnrotation}
  Q(\D\circ\P_2)\leq\min{\{Q(\D),Q(\P_2)\}}=0.
\end{equation}
Thus, reliable quantum transmission is impossible.

The physics behind the possibility of increasing the channel capacity is the deformation of the wave packet due to the relativistic aberration of light. Loosely speaking, the boost reduces the relative width of the wake packet in the transverse plane, which in turn allows for enhanced communication rates. Note that the transverse component of the polarization vector can be controlled via polarizing plates; it is the longitudinal component of the polarization vector which is not controllable by the observers. However, as we have shown, its detrimental effect on information transmission can be reduced by an appropriate Lorentz transformation.

Apart from rigorously quantifying the rate at which two inertial observers can quantum  communicate, our work also sheds light on the intriguing observation made in~\cite{PT} on the presence of non-completely positive (non-CP) dynamics in relativistic transformations of photonic wave packets. Here we conclude that it is more of a mathematical curiosity than having a real physical impact. To see this, we summarize the situation studied in this paper in the diagram in Fig.~\ref{fig:diagram}. We first map a logical qubit $\psi$ to a realistic photonic wave packet $\Psi_0$. We  either may decide to detect the wave packet using the detection given by Eq.~(\ref{eq:W}) (downward arrow labeled $W$) and obtain a helicity density matrix $\vr_0$ or we can Lorentz boost $\Psi_0$ to obtain $\Psi_\zeta=U(\La)\Psi_0$ (wavy line) and then detect. This yields a density matrix $\vr_\zeta$. It may indeed happen that the dotted line connecting $\vr_0$ and $\vr_\zeta$ represents a non-CP map (depends on the boost direction), but the important point is that once $\vr_0$ is received, $\vr_\zeta$ can't be obtained by a Lorentz boost, and vice versa. The relativistic protocol ends by obtaining a helicity density matrix -- only wave packets are Lorentz transformed (that is, transmitted). The Pauli channels $\P'$ and  $\P$ are mutually exclusive.

\subsection*{Acknowledgements}
This work was partially supported by DGAPA-UNAM (under Project No. IN101614). E.C-R thanks CONACyT-Mexico for financial support.

\appendix

\section{Wave packet and output density matrix construction}
\label{ap:wavpackconstr}
\numberwithin{equation}{section}

The degrees of freedom suitable for information transmission in free space are the helicity (circular polarization) states. The reason  we prefer helicity to horizontal/vertical polarization is its conceptual clarity: the helicity is a Poincar\'e invariant~\cite{sexl,tung}. The two Casimir operators of the Poincar\'e group are the squares of the four-momentum operator $P^\mu$ and the Pauli-Lubanski vector $W^\mu$. However, their eigenvalues are zero for massless fields and do not serve as `good' quantum numbers. This is because, based on physical grounds, we take only the $SO(2)$ subgroup of the little group generated by $W^\mu$. Instead, we label the states carrying this particular representation as $\ket{k,\la}$ by the eigenvalues of $P^\mu$ and $W^\mu$ themselves (note that $P^\mu\ket{k,\la}=k^\mu\ket{k,\la}$ and $W^\mu\ket{k,\la}=\la k^\mu\ket{k,\la}$~\cite{sexl,tung}, where $k=(k^0,\bk)$.  The momentum/helicity eigenstates  $\ket{k,\la}$ satisfy the standard normalization condition $\brk{k,\la}{k',\la'}=(2\pi)^3(2k^0)\delta_{\lambda \lambda'} \delta^3(\bk-\bk')$.

The ket notation for the eigenstates suggests that they are elements of a Hilbert space but that is not really the case. The explicit realization of a separable Hilbert space we are interested in is the space of square-integrable functions. But a momentum/helicity eigenstate is not square integrable and real-world physical processes do not generate such states. Realistic photonic states are wave packets whose spatial localization is provided by a Fourier transformation of a square-integrable momentum envelope function $f_\la(\bk)$. The general form of such a state is
\begin{equation}\label{eq:wavepack}
\vert \Psi \rangle = \sum_{\la=\pm} \int f_\lambda(\bk)\vert k, \lambda\rangle \dx{\mu}(k),
\end{equation}
where $\mathrm{d}\mu(k)$ is the relativistic volume element
\begin{equation}\label{eq:CartVolElmnt}
\dx{\mu}(k) = {1\over(2\pi)^3}{1\over 2k^0} \dx^3{k}.
\end{equation}
A Lorentz transformation $\La$ of wave packet~(\ref{eq:wavepack}) induces a unitary transformation $U(\La)$:
\begin{align}
U(\Lambda)\vert \Psi \rangle =& \sum_{\la=\pm} \int  f_\lambda(\bk)U(\Lambda) \vert k, \lambda\rangle  \dx{\mu}(k)\nn\\
=& \,\sum_{\la=\pm} \int  e^{i\la\t_W(k,\La)}  f_\lambda(\bk) \vert \Lambda k, \lambda\rangle  \dx{\mu}(k)\nonumber \\
=& \, \sum_{\la=\pm} \int e^{i\la\t_W(k,\La)} f_\lambda(\Lambda^{-1}\bk) \vert  k, \lambda\rangle \dx{\mu}(k),
\end{align}
where $\t_W(k,\La)$ is Wigner's angle~\cite{wigner}, whose explicit form can be found in~\cite{Gingrich,caban}. For the  special case of a boost in the wave packet propagation direction studied in this paper the phase is zero. This can be explicitly shown as follows.

The little group for massless particles is the Euclidean group in two dimensions $E_2$, which is a semi direct product of the rotation group in two dimensions $SO(2)$, and the group of translations in the plane $T_2$. A general element $W \in E_2$ can be written as $W = TR$, where $T\in T_2$ and $R \in SO(2)$ \cite{tung}. A nontrivial representation of the translation group in the Hilbert space of massless particles yields particle states labeled by continuous internal degrees of freedom \cite{Weinberg}. Since no such particles are known to exist, the group $T_2$ is represented trivially and only the rotation part of the little-group element plays a role in the transformation rule for the massless case. In particular, one-particle photonic states transform as
\begin{equation}
U(\Lambda)\vert p, \lambda\rangle = \mathrm{e}^{\mathrm{i}\lambda \vartheta(\Lambda, p)} \vert p, \lambda\rangle,
\end{equation}
where $\lambda = \pm 1$. The labels $p$ and $\lambda$ denote, respectively, the four-momentum and the helicity of the photon.

We now show that the Wigner angle $\vartheta(\Lambda, p)$ vanishes when $\Lambda$ is a pure boost along the $z$-axis with velocity $v = \tanh \zeta$, with $\zeta \in \mathbb{R}$, and $p = \omega(1,\cos\phi \sin\theta, \sin\phi \sin\theta, \cos\theta)^T$ is an arbitrary (null) four-momentum vector. This is equivalent to showing that the little-group element
\begin{equation}
W(\Lambda, p) = L^{-1}_{\Lambda p}\Lambda L_p
\end{equation}
is a pure translation for this choice of $\Lambda$.

The Lorentz transformation $L_p$ takes the standard four-vector $k = (1,0,0,1)^T$ to $p$, and is defined by
\begin{equation}
L_p = R(\hat{p})B_z(\xi),
\end{equation}
where $B_z(\xi)$ is a pure boost along the $z$-axis, with rapidity $\xi = -\ln \omega$, which takes the four-vector $k$ to the four-vector $(\omega,0,0,\omega)^T$, and $R(\hat{p})$ is a rotation that takes the latter to $p$. The rotation $R(\hat{p})$ is defined as $R(\hat{p}) = R_z(\phi)R_y(\theta)$, where $R_z(\phi)$ is a rotation along the $z$ axis by an angle $\phi$, and $R_y(\theta)$ is a rotation along the $y$ axis by an angle $\theta$.

On the other hand, the transformed four-vector $p$ is given by
\begin{equation}
\Lambda p = \omega \begin{bmatrix}
\cosh\zeta-\sinh\zeta\cos\theta \\ \cos\phi\sin\theta \\ \sin\phi\sin\theta \\ -\sinh\zeta+\cosh\zeta\cos\theta
\end{bmatrix},
\end{equation}
so that the transformation $L_{\Lambda p}$ reads $L_{\Lambda p} = R_z(\tilde{\phi})R_y(\tilde{\theta})B_z(\tilde{\xi})$, where $\tilde{\phi} = \phi$ (a boost along $z$ does not affect the azimuthal angle), $\tilde{\theta} = \mathrm{arcsin}\left[(\omega\cosh\zeta-\omega\sinh\zeta\cos\theta)^{-1}\sin\theta\right]$ and $\tilde\xi = -\ln(\omega\cosh\zeta-\omega\sinh\zeta\cos\theta)$. Putting all the pieces together, we find
\begin{align}
W(\Lambda, p) =& L^{-1}_{\Lambda p}\Lambda L_p \nonumber \\
=& B_z(-\tilde{\xi})R_y(-\tilde{\theta})R_z(\phi)B_z(\zeta)R_z(\phi)R_y(\theta)B_z(\xi) \nonumber \\
=& B_z(-\tilde{\xi})R_y(-\tilde{\theta})B_z(\zeta)R_y(\theta)B_z(\xi),
\end{align}
where we have used the fact that rotations and boosts along the same axis commute. At this point it is clear that the little group element $W(\Lambda, p)$ has no contribution from $SO(2)$ since the rotations along $z$ have canceled out. Indeed, a direct calculation shows
\begin{equation}
\label{t2}
W(\Lambda, p) =
\begin{bmatrix}
1+ \frac{1}{2} \vec{a}^2 & a_1 & a_2 & -\frac{1}{2} \vec{a}^2 \\
a_1 & 1 & 0 & -a_1 \\
a_2 & 0 & 1 & -a_2 \\
\frac{1}{2} \vec{a}^2 & a_1 & a_2 &  1- \frac{1}{2} \vec{a}^2
\end{bmatrix},
\end{equation}
where $\vec{a} = (a_1, a_2)^T$, with $a_1 = (\mathrm{coth}\zeta-\cos\theta)^{-1}\mathrm{e}^{\xi}\sin\theta$, and $a_2=0$. This has the form of a pure translation by the vector $\vec{a}$ in the $x-y$-plane \cite{Duncan}. Therefore, we have $\vartheta(\Lambda, p) = 0$ as was to be shown.

It is no surprise that boosts along the $z$ axis induce a different behavior of the Wigner phase in contrast to boosts in any other direction, since the standard vector $k$ is defined so that its spatial part points in the $z$ direction. We have here chosen the $z$ axis to be the main direction of propagation of the wave packet for calculational convenience only. Any other choice of this direction would of course yield the same results for appropriate definitions of standard vector $k$ and standard boosts $L_p$.

It is natural for the momentum distribution of a realistic wave packet to possess axial symmetry. One such choice is a Gaussian profile whose form reads
\begin{equation}\label{eq:wavepackGauss}
f({\bk})={1\over\s_z\s^2(2\pi)^{3/2}}\exp{\left(-{k_1^2+k_2^2\over2\s^2}\right)\exp{\left(-{(k_3-k_p)^2\over2\s^2_z}\right)}},
\end{equation}
where $k_p>0$ is the mean value determining the average direction of wave packet propagation. This function is normalized in the following sense:
$$
\int_{-\infty}^\infty f(\bk)\dx^3{k}=1.
$$
But this is not a covariant normalization. For the case of wave packets used in relativistic situations (meaning that at least one of the observers is moving at a relativistic speed),  we are interested in the following condition being satisfied:
\begin{equation}\label{eq:covNorm}
\int_{-\infty}^\infty  |f(\mathbf{k})|^2 \dx{\mu}(k) =1,
\end{equation}
where we have redefined the envelope function
\begin{equation}\label{eq:genWPGauss}
|f(\bk)|^2 = {1\over N'} \exp\left(-\frac{k_1^2+k_2^2}{\s^2}\right)\exp\left(-\frac{(k_3-k_p)^2}{\s_z^2}\right).
\end{equation}
Condition~(\ref{eq:covNorm}) ensures that the overall probability is conserved for a Lorentz transformed  wave packet. There is no need to find $N'$ since we will make a certain physically motivated approximation.
In particular, we will assume that the variance of the distribution in the $z$ direction is much smaller than the variance in the radial direction, i.e., $\sigma_z/\sigma \ll 1$. In the end we have to work with the approximated wave packet in spherical coordinates. So before we make the approximation, we transform Eq.~(\ref{eq:covNorm}) to the desired coordinate system. The four-vector $k^\mu=(k^0,k_1,k_2,k_3)$ becomes $k^\mu=k^0(1,\sin{\t}\cos{\phi},\sin{\t}\sin{\phi},\cos{\t})$ and after a Lorentz boost has been applied we obtain $\tilde{k}^\mu=(\Lambda^{-1})^\mu_{ \ \nu}k^\nu$ where
\begin{equation}\label{eq:boost}
\Lambda^{-1}\equiv B_z^{-1}(\zeta)
=\begin{bmatrix}
\cosh{\zeta} && 0 && 0 &&  \sinh{\zeta} \\
0 && 1 && 0 && 0 \\
0 && 0&& 1 && 0 \\
 \sinh{\zeta} && 0 && 0 &&   \cosh{\zeta}
\end{bmatrix},
\end{equation}
with $\zeta=\atanh{v_z}$ being the rapidity and $v_z$ the speed. Hence
\begin{equation}\label{eq:ktilde}
\tilde{k}^\mu
    =k^0
    \begin{bmatrix}
    \cosh{\zeta}  + \sinh{\zeta}\cos{\t} \\
    \sin{\t}\cos{\phi} \\
    \sin{\t}\sin{\phi} \\
    \sinh{\zeta} + \cosh{\zeta}\cos{\t}
    \end{bmatrix}.
\end{equation}
and the Cartesian volume element becomes a spherical `relativistic' volume element
\begin{equation}\label{eq:wedgeForm}
  \dx{k_1}\wedge\dx{k_2}\wedge\dx{k_3}=(k^0)^2\sin{\t}(\cosh{\zeta}+\sinh{\zeta}\cos{\t})\dx{k^0}\wedge\dx{\t}\wedge\dx{\phi},
\end{equation}
where $\wedge$ stands for the wedge product. Consequently,  Eq.~(\ref{eq:CartVolElmnt}) transforms into
\begin{equation}\label{eq:SpherVolElmnt}
  \dx{\mu}(k) = {1\over(2\pi)^3}{k^0\sin{\t}\over2}\dx{k^0}\dx{\t}\dx{\phi}
\end{equation}
and Eq.~(\ref{eq:genWPGauss}) becomes
\begin{equation}\label{eq:genWPSpher}
  |f(\La^{-1}\bk)|^2 = {1\over N'} \exp\left(-\frac{(k^0)^2\sin^2{\t}}{\s^2}\right)\exp\left(-\frac{(k^0(\sinh{\zeta} + \cosh{\zeta}\cos{\t})-k_p)^2}{\s_z^2}\right).
\end{equation}
At this point we introduce the aforementioned approximation and set
\begin{subequations}
\begin{align}
|f(\La^{-1}\bk)|^2
& = {1\over N} \exp\left(-\frac{(k^0)^2\sin^2{\t}}{\s^2}\right)\delta( k^0(\sinh{\zeta}+\cosh{\zeta}\cos{\t})-k_p) \\
& = {1\over N} \exp\left(-\frac{(k^0)^2\sin^2{\t}}{\s^2}\right)\delta\left(k^0-\frac{k_p}{\sinh{\zeta}+\cosh{\zeta}\cos{\t}}\right)
{k_p\over\sinh{\zeta}+\cosh{\zeta}\cos{\t}},
\end{align}
\end{subequations}
where we have used the delta function identity: $\delta(ax) = \delta(x)/|a|$. The integral over $k^0$ yields
\begin{equation}\label{eq:approxEnvelope}
|f(\La^{-1}\bk)|^2 = {1\over N} \exp\left(-\frac{ \sin^2\t}{\Gamma^2(\sinh\zeta+\cosh\zeta\cos\t)^2 }\right) \frac{1}{\sinh{\zeta}+\cosh{\zeta}\cos{\t}},
\end{equation}
where $\Gamma = \s/k_p$. We find the normalization condition by evaluating the complete integral
\begin{align}\label{eq:covNormSpherInt}
  &N\int|f(\La^{-1}\bk)|^2\dx{\mu}(k)  =\nn\\
   &{k_p\over2(2\pi)^3}\int_0^{\t_c}\int_0^{2\pi}\exp{\left(-\frac{\sin^2\t}{\Gamma^2(\sinh\zeta+\cosh\zeta\cos\t)^2}\right)}
  {\sin{\t}\over(\sinh{\zeta}+\cosh{\zeta}\cos{\t})^2}\dx{\t}\dx{\phi},
\end{align}
where we have used Eqs.~(\ref{eq:SpherVolElmnt}) and (\ref{eq:approxEnvelope}). We further define the kernel $K(\t,\zeta,\Gamma)$ to be
\begin{equation}\label{eq:kernel}
  K(\t,\zeta,\Gamma)\df\exp{\left(-\frac{\sin^2\t}{\Gamma^2(\sinh\zeta+\cosh\zeta\cos\t)^2}\right)}
  {\sin{\t}\over(\sinh{\zeta}+\cosh{\zeta}\cos{\t})^2}.
\end{equation}

There are two covariant options for the upper bound angle $\t_c$. Either it can be $\t_c=\pi$, which comes directly from the change of variables given by the transformation in Eq.~(\ref{eq:ktilde}), or we can set
\begin{equation}\label{eq:aberrationAngle}
\t_c = \arccos{\left(-\tanh{\zeta}\right)}.
\end{equation}
In both cases the integral is relativistically invariant since the normalization is given by
\begin{equation}\label{eq:WPNorm1}
N={k_p\over(2\pi)^3}\Gamma\pi^{3\over2}\exp{\Big(-{1\over\Gamma^2}\Big)}\Big(1-\erf{1\over\Gamma}\Big)
\end{equation}
for $\t_c=\pi$ and
\begin{equation}\label{eq:WPNorm2}
N={k_p\over2(2\pi)^3}\Gamma\pi^{3\over2}\exp{\Big(-{1\over\Gamma^2}\Big)}\Big(1-\erf{1\over\Gamma}\Big)
\end{equation}
for $\t_c=\arccos{\left(-\tanh{\zeta}\right)}$. A closer analysis of Eq.~(\ref{eq:covNormSpherInt}) reveals that the approximation leads to two Gaussian wave packets (peaked at $\pm k_p$ in the $z$ momentum component) and therefore moving in the opposite direction. We are, however, interested in only the one with a positive $z$ component of the momentum. Hence, the second option for $\t_c$ corresponds to the physically interesting situation of just a single-direction traveling wave packet. Of course, the other option would be to define the envelope function  to be identically zero in the region corresponding to $k_p<0$ (before applying a Lorentz transformation) and then we could set $\t_c=\pi$.

The second angle (Eq.~(\ref{eq:aberrationAngle})) can be derived from the usual formulas for the relativistic aberration of light given by how a polar angle $\tilde\t$ of one observer is perceived by another inertial observer as a function of $\zeta$:
\begin{equation}
\tilde\t = \arctan\left(\frac{\sin{\t}}{\sinh{\zeta}+\cosh{\zeta}\cos{\t}}\right).
\end{equation}
By setting $\tilde\t=\pi/2$ we obtain the angle $\t_c$ in Eq.~(\ref{eq:aberrationAngle}).

By plugging Eqs~(2),~(4) and~(\ref{eq:approxEnvelope}) into Eq.~(3) we obtain the explicit expressions of the density matrix components of Eq.~(5):
\begin{subequations}\label{eq:rhoOutComps}
\begin{align}
  \vr_{\zeta,00} & = {1\over N}\int_0^{2\pi}\int_0^{\t_c} K(\t,\zeta,\Gamma){g_1+g_2\cos{\chi}\sin{\xi}\over1-\cos^2{\phi}\sin^2{\t}}\dx{\phi}\dx{\t},\\
  \vr_{\zeta,11} & = {1\over N}\int_0^{2\pi}\int_0^{\t_c} K(\t,\zeta,\Gamma){g_3+g_4\cos{\chi}\sin{\xi}\over1-\sin^2{\phi}\sin^2{\t}}\dx{\phi}\dx{\t},\\
  \vr_{\zeta,01} & = {1\over N}\int_0^{2\pi}\int_0^{\t_c} K(\t,\zeta,\Gamma){g_5\sin{\chi}\sin{\xi}+i g_6\cos{\xi}\over\sqrt{1-\cos^2{\phi}\sin^2{\t}}\sqrt{1-\sin^2{\phi}\sin^2{\t}}}\dx{\phi}\dx{\t},
\end{align}
\end{subequations}
where
\begin{subequations}\label{eq:gFcns}
\begin{align}
  g_1 & = {1\over2}(\cos^2{\phi}\cos^2{\t}+\sin^2{\phi}),  \\
  g_2 & = {1\over2}(\cos^2{\phi}\cos{2\phi}\cos^2{\t}-\cos{2\phi}\sin^2{\phi}+\cos{\t}\sin^2{2\phi}),\\
  g_3 & = {1\over2}(\sin^2{\phi}\cos^2{\t}+\cos^2{\phi}),\\
  g_4 & = {1\over2}(\sin^2{\phi}\cos{2\phi}\cos^2{\t}-\cos{2\phi}\cos^2{\phi}-\cos{\t}\sin^2{2\phi}),\\
  g_5 & = {1\over4}(2\cos^2{2\phi}\cos{\t}+\sin^2{2\phi}+\cos^2{\t}\sin^2{2\phi}),\\
  g_6 & = -{1\over2}\cos{\t}.
\end{align}
\end{subequations}
We can easily read off the $\la_i$ parameters responsible for the Pauli channel output structure Eq.~(5).
\begin{subequations}\label{eq:lamdaIntegrals}
  \begin{align}
    \la_1 & = {2\over N}\int_0^{2\pi}\int_0^{\t_c} K(\t,\zeta,\Gamma){g_5\over\sqrt{1-\cos^2{\phi}\sin^2{\t}}\sqrt{1-\sin^2{\phi}\sin^2{\t}}}\dx{\phi}\dx{\t},\\
    \la_2 & = {2\over N}\int_0^{2\pi}\int_0^{\t_c} K(\t,\zeta,\Gamma){g_6\over\sqrt{1-\cos^2{\phi}\sin^2{\t}}\sqrt{1-\sin^2{\phi}\sin^2{\t}}}\dx{\phi}\dx{\t},\\
    \la_3 & = {2\over N}\int_0^{2\pi}\int_0^{\t_c} K(\t,\zeta,\Gamma){g_2\over1-\cos^2{\phi}\sin^2{\t}}\dx{\phi}\dx{\t}.
  \end{align}
\end{subequations}
For $\la_i$ to be the Pauli channel parameters one has to show that condition~(6) is satisfied. Indeed this is true. From the form of $g_j$ functions in Eqs.~(\ref{eq:gFcns}), one can see that for $j=1$ we get
\begin{equation}\label{eq:provePauli}
{g_1\over1-\cos^2{\phi}\sin^2{\t}}={g_3\over1-\sin^2{\phi}\sin^2{\t}}={1\over2}.
\end{equation}
For $j=2$ the absolute values of the integrands in~(6) are not equal but, rather, shifted by $\pi/2$. This can be seen by setting $\cos{\phi}=\sin{\phi'}$ (hence $\phi'=\phi+\pi/2$ and $\dx{\phi}=\dx{\phi'}$), and therefore $\cos{2\phi}=-\cos{2\phi'}$ and $\sin{2\phi}=-\sin{2\phi'}$. By plugging these expressions into $g_2$ we obtain $-g_4$ and prove identity~(6) for all $\zeta\in\bbR$ and arbitrary $\Gamma$.

\section{Detailed derivation of the Pauli channel parameters $\la_i$ for $\zeta=0$}
\label{ap:detailedcalc}

We derive analytic and perturbative expressions for the coefficients $\la_i$ of the Pauli channel for the observer at rest where $\zeta=0$. The derivation of the Lorentz boosted Pauli channel follows a similar path but the complexity of the calculations is far higher since the components of the output density matrix themselves are not relativistically invariant. The components of the density matrix are calculated from Eq.~(3) by using (4) and the kernel, Eq.~(\ref{eq:kernel}). We omit the omnipresent constant ${k_p/(2\pi)^3}$ and start with the output density matrix normalization
\begin{equation}\label{eq:trace}
  N=\Tr{}[\vr]=\int_0^\infty{\exp{(-{s\over\Gamma^2})}\over2\sqrt{1+s}}\dx{s}
  =\Gamma\pi^{3\over2}\exp{\Big(-{1\over\Gamma^2}\Big)}\Big(1-\erf{1\over\Gamma}\Big).
\end{equation}
This is an entirely independent confirmation of the wave packet normalization Eq.~(\ref{eq:WPNorm2}).
For $\zeta=0$ we obtain from Eqs.~(\ref{eq:lamdaIntegrals})
\begin{subequations}\label{eq:lamdaIntegralszeta0}
  \begin{align}
    \la_1 & = {2\over N}\int_0^{2\pi}\int_0^{\pi/2} e^{-\tan{\t}^2/\Gamma^2}{\sin{\t}\over\cos^2{\t}}{g_5\over\sqrt{1-\cos^2{\phi}\sin^2{\t}}\sqrt{1-\sin^2{\phi}\sin^2{\t}}}\dx{\phi}\dx{\t},\\
    \la_2 & = {2\over N}\int_0^{2\pi}\int_0^{\pi/2} e^{-\tan{\t}^2/\Gamma^2}{\sin{\t}\over\cos^2{\t}}{g_6\over\sqrt{1-\cos^2{\phi}\sin^2{\t}}\sqrt{1-\sin^2{\phi}\sin^2{\t}}}\dx{\phi}\dx{\t},\\
    \la_3 & = {2\over N}\int_0^{2\pi}\int_0^{\pi/2} e^{-\tan{\t}^2/\Gamma^2}{\sin{\t}\over\cos^2{\t}}{g_2\over1-\cos^2{\phi}\sin^2{\t}}\dx{\phi}\dx{\t}.
  \end{align}
\end{subequations}
We start with $\la_3$, whose form can be obtained analytically. By integrating over $\phi$ and using the substitution $s=\tan^2{\t}$, we get
$$
\la_3={2\pi\over N} \int_0^\infty\exp{\Big(-{s\over\Gamma^2}\Big)}{2+s-2\sqrt{1+s}\over s^2}\dx{s}.
$$
This has the form of a Laplace transform (denoted $\L$) but the double pole at $s=0$ is troublesome. To get rid of it we introduce an auxiliary variable, $p=1/\Gamma^2$, and twice differentiate the integrand with respect to it, leading to $f(s)=2+s-2\sqrt{1+s}$. The rest is a routine calculation provided by Mathematica:
$$
\L(f(s))(p)={1\over p^2}\big(1-e^p\sqrt{2\pi}\erf{\sqrt{p}}\big).
$$

We reverse the derivatives by two antiderivatives
$$
\la_3={2\pi\over N}\int\dx{p}\int\dx{p}\left({1\over p^2}\big(1-e^p\sqrt{2\pi}\erf{\sqrt{p}}\big)\right)+c,
$$

\noindent where $c$ is an integration constant given by $\lim_{p\to\infty}\la_3=0$ because of an exponential tail.  We find that
\begin{equation}\label{eq:lambda3}
\lambda_3={4\pi\over3N}\left({2p^2}\,\pFq{2}{2}{1,1}{{5\over2},3}{p}+3\left(\pi i(2p-1)\erf{i\sqrt{p}}+2\sqrt{\pi p}\exp{p}-\log{p}+2p\big(\g-3+\log{4p}\big)\right)\right)+c,
\end{equation}

\noindent where $_p F_q$ is a generalized hypergeometric function, $\g$ is the Euler-Mascheroni constant and
$$
c=-2\pi(\g+1+\log{4}).
$$

For $\la_1$ and $\la_2$, a different strategy has to be used. We  illustrate it in the calculation of $\la_1$. Using the same substitution as before and after integrating over $\phi$, we obtain
\begin{align}\label{eq:lambda1}
  \la_1&={2\over N}\int_0^\infty\exp{\Big(-{s\over\Gamma^2}\Big)}\times\nn\\
  &\Bigg(q_1E\bigg[-{s^2\over4(1+s)}\bigg]+q_2K\bigg[-{s^2\over4(1+s)}\bigg]  +q_3E\bigg[{s^2\over(2+s)^2}\bigg] +  q_4K\bigg[{s^2\over(2+s)^2}\bigg]        \Bigg)\dx{s},
\end{align}
where $K[z]$ and $E[z]$ are the complete elliptic integrals of the first and second kind, respectively, and $q_i$ are polynomials
\begin{align}\label{eq:qPolys}
  q_1 & =-\frac{2 \sqrt{s+1}}{s^2}+\frac{2}{s^2 (s+1)}+\frac{3}{s (s+1)}+\frac{1}{s+1}, \\
  q_2 & = \frac{2}{s^2 \sqrt{s+1}}-\frac{2}{s^2 (s+1)}+\frac{2}{s
   \sqrt{s+1}}+\frac{1}{2 \sqrt{s+1}}-\frac{3}{s (s+1)}-\frac{1}{s+1},\\
  q_3 & = -\frac{2}{s^2}+\frac{2}{s^2 \sqrt{s+1}}-\frac{1}{s}+\frac{2}{s
   \sqrt{s+1}}+\frac{1}{2 \sqrt{s+1}},\\
  q_4 & = \frac{2}{s^2}-\frac{2}{s^2 \sqrt{s+1}}+\frac{1}{s}-\frac{2}{s \sqrt{s+1}}.
\end{align}
To get rid of singularities the derivative trick would work here as well, but unfortunately, after this step, we do not know how to evaluate the Laplace transform of the elliptic functions with the arguments we have. Instead, we resolve the integral in a certain asymptotic manner and express the result perturbatively in $1/\Gamma^2$. For this purpose we realize that we may take
\begin{align}
  \la^L_1&={2\over N}\int_0^L\exp{\Big(-{s\over\Gamma^2}\Big)}\times\nn\\
  &\Bigg(q_1E\bigg[-{s^2\over4(1+s)}\bigg]+q_2K\bigg[-{s^2\over4(1+s)}\bigg]  +q_3E\bigg[{s^2\over(2+s)^2}\bigg] +  q_4K\bigg[{s^2\over(2+s)^2}\bigg]        \Bigg)\dx{s},
\end{align}
where $L$ is high enough to approximate the integral as close as one wishes. Then, we may expand the exponential function in $s$ around zero and are allowed to exchange the sum and integral
\begin{align}
  \tilde\la^L_1&={2\over N}\sum_{n=0}^\infty{1\over\Gamma^{2n}}\kappa_n,
\end{align}
where
\begin{align}
\kappa_n&={(-)^n\over n!}\int_0^L\,s^n\Bigg(q_1E\bigg[-{s^2\over4(1+s)}\bigg]+q_2K\bigg[-{s^2\over4(1+s)}\bigg]  +q_3E\bigg[{s^2\over(2+s)^2}\bigg] +  q_4K\bigg[{s^2\over(2+s)^2}\bigg]        \Bigg)\dx{s}.
\end{align}
The solution of the $\kappa_n$ integrals is unknown to the authors either but the important point is that they are mere coefficients of the $\Gamma$ expansion. Hence they do no depend on the channel parameter and once they are calculated (numerically or otherwise) they are valid for an arbitrary Lorentz boost.

Similarly, we obtain
\begin{align}\label{eq:lambda2}
  \la_2&={2\over N}\int_0^\infty\exp{\Big({s\over\Gamma^2}\Big)}{1\over2}\Bigg({1\over\sqrt{1+s}}K\bigg[-{s^2\over4(1+s)}\bigg]
  -{1\over2+s}K\bigg[{s^2\over(2+s)^2}\bigg]\Bigg)\dx{s}
\end{align}
and define
\begin{equation}
   \tilde\la^L_2={2\over N}\sum_{n=0}^\infty{1\over\Gamma^{2n}}\iota_n,
\end{equation}
where
\begin{align}
  \iota_n  = {(-)^n\over n!}\int_0^L\,s^n{1\over2}\Bigg({1\over\sqrt{1+s}}K\bigg[-{s^2\over4(1+s)}\bigg]
  +{1\over2+s}K\bigg[{s^2\over(2+s)^2}\bigg]\Bigg)\dx{s}.
\end{align}

\end{document}